\documentclass[12pt,preprint]{aastex}
\usepackage[cp1251]{inputenc}
\usepackage[english]{babel}
\usepackage{amssymb}
\usepackage{amsmath}

\shorttitle{Simulation of the interaction between WASP-12b and its host star}
\shortauthors{Bisikalo et al.}

\newcommand{\lp}[1]{\mathop{\rm{L_#1}}}

\begin{document}

\title{3D gas dynamic simulation of the interaction between the exoplanet
WASP-12b and its host star}

\author{D.~Bisikalo, P.~Kaygorodov, D.~Ionov, V.~Shematovich}
\affil{Institute of Astronomy, Russian Academy of Sciences, Moscow, Russia}
\email{bisikalo@inasan.ru}

\author{H.~Lammer}
\affil{Space Research Institute, Austrian Academy of Sciences, Graz, Austria}

\and

\author{L.~Fossati}
\affil{Argelander Institut f\"ur Astronomie der Universit\"{a}t Bonn, Bonn, Germany}

\begin{abstract}
HST transit observations in the near-UV performed in 2009 made
WASP-12b one of the most ``mysterious'' exoplanets; the system
presents an early-ingress, which can be explained by the presence
of optically thick matter located ahead of the planet at a
distance of $\sim$\,4--5 planet radii. This work follows previous
attempts to explain this asymmetry with an exospheric outflow or a
bow shock, induced by a planetary magnetic field, and provides a
numerical solution of the early-ingress, though we did not
perform any radiative transfer calculation. We performed pure 3D gas dynamic
simulations of the plasma interaction between WASP-12b and its
host star, and describe the flow pattern in the system. In
particular, we show that the overfilling of the planet's Roche
lobe leads to a noticeable outflow from the upper atmosphere in
the direction of the $\lp1$ and $\lp2$ points. Due to the
conservation of the angular momentum, the flow to the $\lp1$ point
is deflected in the direction of the planet's orbital motion,
while the flow towards $\lp2$ is deflected in the opposite
direction, resulting in a non-axisymmetric envelope, surrounding
the planet. The supersonic motion of the planet inside the stellar
wind leads to the formation of a bow shock with a complex shape.
The existence of the bow shock slows down the outflow through the
$\lp1$ and $\lp2$ points, allowing us to consider a long-living
flow structure which is in the steady-state.
\end{abstract}

\maketitle

\section{Introduction}
WASP-12 is a late F-type main sequence star
\citep[$M_*=1.35\,M_\odot$, $R_*=1.6\,R_\odot$, $T_{\rm eff}=6250\pm100\,K$, log $g=4.2\pm0.2$;][]{Fossati-et-al:2010b}
with a magnitude of $V\sim\,11^m.6$. The star is at a distance of
about 400\,pc from the Sun \citep{Fossati-et-al:2010b}. The star
hosts a transiting ``hot-Jupiter'', WASP-12b, with a mass of
$M_{\rm pl}=1.41\pm0.1\,M_{\rm Jup}$ and radius of $R_{\rm
pl}=1.74\pm0.09\,R_{\rm Jup}$ \citep{Chan-et-al:2011}. WASP-12b
revolves in a rather circular orbit \citep{Campo-et-al:2011} with
a period of $\sim$\,1.09\,days \citep{Hebb-et-al:2009} at a
distance of 0.0229\,AU ($\sim$\,3 stellar radii) from its host
star.

WASP-12b was observed in 2009 in the near-UV with the Cosmic
Origins Spectrograph (COS) on board of HST
\citep{Fossati-et-al:2010a}. The observed spectral range was
2540--2810\AA, divided into three bands of 40\,\AA\ each. Analysis
of the COS light curves showed that the transit, in two of the
three observed spectral regions, is considerably deeper than in
the visible. Besides, an early ingress was also discovered. The
HST data show that in the near-UV the transit starts approximately
50 minutes earlier than in the visible, meaning that ahead of the
planet, at a distance of 4--5 planet radii, there is a region of
absorbing matter.

So far, two explanations for this optically thick matter ahead of
the planet have been proposed. The first explanation is based on
mass transfer from the planet to the star \citep{Lai-et-al:2010,
Li-et-al:2010}. Indeed, the distance of the center of the planet
to the $\lp1$ point is only $\sim\,1.85\,R_{\rm pl}$, and its
upper atmosphere exceeds the Roche lobe.

The size of the collision dominated atmosphere is limited by the
exobase level $R_{\rm ex}$, where the mean free path is equal to
the atmosphere's scale height. For WASP-12b's atmosphere, which
has an expected density at the base of the thermosphere of
$\rho_{\rm pl}=2.7\times10^{-14}\,\text{g}/\text{cm}^{3}$ {(see
Sect.~\ref{model}) and is heated by the stellar EUV radiation to a
temperature of $T_{\rm pl}=10^4$\,K, the exobase would expand
beyond the Roche lobe distance. Using standard physics of binary
stars \citep[see e.g.,][]{Pringle:1985, Savonije:1979}, to
estimate how much the upper atmosphere overfills the Roche lobe,
one can calculate the volume Roche radius $R_{\lp1}$, obtaining
$\sim\,1.37\,R_{\rm pl}$ and the degree of overfilling $\Delta
R/R_{\lp1}=(R_{\rm ex}-R_{\lp1})/R_{\lp1}\approx0.13$. Such a
large degree of overfilling leads to a powerful outflow of
material through the $\lp1$ and $\lp2$ points. The natural result
of this outflow is the formation of an accretion disk surrounding
the star, as for close binaries, where the flow structure is well
studied, both analytically \citep[see, e.g.,][]{Lubow-Shu:75} and
numerically \citep[see,
e.g.,][]{Bisikalo-et-al:2003,Bisikalo:2008,Zhilkin:2012}.

If the accretion disk is axisymmetric it does not produce any
effect on the transit shape. However, in the region where the
stream interacts with the disk an observable extended shock wave
forms \citep{Boyarchuk-et-al:2002, Bisikalo-et-al:2003}. This
hypothesis was used by \citet{Lai-et-al:2010}, who supposed that
the stream from $\lp1$ itself and/or the region of the stream-disk
interaction can cause the early ingress. Although the idea is
reasonable, a more detailed analysis shows that, with such a large
degree of Roche lobe overfilling, the lifetime of the planet
atmosphere would be very short. In accordance with equations used
in the theory of close binaries \citep[see e.g.,][]{Pringle:1985},
the mass loss rate is defined as:
\begin{equation}
\dot{M}/M=(\Delta R/R)^3 \sqrt{\frac{GM}{R^3}} \,.
\label{f1}
\end{equation}
\noindent Using this equation and assuming a total mass of the
WASP-12b's atmosphere of $1.41 M_{\rm Jup}$, we find that the
lifetime of the atmosphere is no longer than several years.
Accounting for the adjustment of the mass loss rate with time, one
can then more accurately estimate the lifetime of the outflowing
atmosphere. Let us assume that the temperature of the thermosphere
is constant, which is a reasonable assumption as WASP-12 is an
inactive star \citep{Fossati-et-al:2010b}. In this case the mass
decrease of a gas giant planet leads to an  increase of the Roche
lobe overfilling, and consequently an increase of the mass loss
rate. As a matter of fact, for a gas giant planet, equation
(\ref{f1}) gives an estimate of the maximum planet life time. On
the other hand, for a planet with a rocky core the mass of
its atmosphere is just a small fraction of the total mass, therefore one
can assume that the size of the Roche lobe will be constant,
independently of the atmosphere loss rate. Hence, the part of the
atmosphere, which lies above the Roche lobe, will flow out till the
atmosphere itself becomes smaller than the Roche lobe, arresting the mass loss.
In this way, for rocky planets one can use equation (\ref{f1})
to obtain a reasonable estimate of the period of time in which the
planet is actively losing mass. In other words, if the overfilling
is high, the lifetime of the atmosphere will be short for any kind
of planet, hence a small chance to observe it. Therefore the
hypothesis proposed by \citet{Li-et-al:2010} and
\citet{Lai-et-al:2010}, most likely does not allow an adequate
explanation of the processes ongoing in the WASP-12 system.
Nevertheless this probability is not zero; it is indeed quite
possible that for systems where the dynamic pressure of the wind
is small the scenario proposed by \citet{Lai-et-al:2010} is
realistic. To some extent our work is a further development of
this idea.

The second possible explanation is the presence of an optically
thick bow shock region ahead of the planet, which possesses a
proper magnetic field \citep{Vidotto-et-al:2010,
Vidotto-et-al:2011}. This interpretation is based on the fact that
the planet orbits with a supersonic velocity inside the stellar
wind, forming a bow shock at a distance of $\sim$\,4--5\,$R_{\rm
pl}$ ahead of the planet. \citet{llama:2011} performed radiative
transfer calculations based on Vidotto et al.'s model, showing that
multiple shock configurations would fit the near-UV light curve,
hence the need of further observations to constrain the model parameters.

\citet{Ionov-et-al:2012} adopted a model similar to that of
\citet{Vidotto-et-al:2010}, but assumed that the planet does not
possess any magnetic field. In this case the bow shock forms
immediately at the level of the upper atmosphere. Although the
authors were able to qualitatively model the light curve, the
shock results to be at a distance significantly different from the
observed one. The distance between the contact discontinuity and
the center of the planet can be defined analytically considering
the dynamic pressure balance from both sides of the contact
discontinuity \citep{Baranov}. For the considered case, this
distance should not be greater than $\sim$\,1.88\,$R_{\rm pl}$ and
the wave front should be located at $\sim$\,2.23\,$R_{\rm pl}$
from its center \citep{Verigin}, half of that indicated by the
observations \citep[see][]{Lai-et-al:2010}. It is important to
note that both bow shock models
\citep{Ionov-et-al:2012,Vidotto-et-al:2010} do not account for the
presence of the outflowing atmosphere. In presence of an
overfilling Roche lobe the balance equations should take
into account the dynamic and thermal pressure of the outflowing
atmospheric gas. In this way, the positions of the bow shock and
contact discontinuity will be defined by the gas properties.
\citet{Vidotto-et-al:2010} showed that the thermal pressure is
negligible for a barometric distribution of the atmospheric gas.
In case instead of an outflowing atmosphere, the gas from the vicinity
of the L$_1$ point will form a stream with a slowly decreasing density,
hence increasing the importance of the thermal pressure. More importantly,
the stream from the L$_1$ point accelerates in the gravitational field 
of the star, increasing its radial velocity as $\sqrt r$. As a consequence, 
the dynamic pressure of the stream will increase ($\sim r$), while the 
importance of the magnetic field will decrease as $r^{-3}$ ($B^2 \sim r^{-6}$). 
This means that for an outflowing atmosphere the role of the stream in 
the solution should be extremely relevant.

The aim of this study is the application of a 3D gas dynamic model
to the gaseous envelope of WASP-12b, in order to provide an
alternative explanation of the observed early ingress. This work
will be followed by a more comprehensive study describing the
sensitivity of the results on the assumption and presenting
comparisons between the model and the available observations. The
work is organized as follows: in Sect. 2 we describe our model and
the assumed input parameters; in Sect. 3 we present the results of
the numerical simulations; and in Sect. 4 the main results are
discussed.

\section{Model description}
\label{model} The atmospheric parameters of WASP-12b are not well
known, therefore we assume an isothermal upper atmosphere with a
temperature, which is defined by the host star's deposition of
XUV-energy, and is similar to that obtained for other
hot-Jupiters: $T_{\rm pl}=10^4$\,K
\citep{Yelle:2004,Koskinen-et-al:2012}. Note that the assumed
temperature of the thermosphere does not correspond to the
equilibrium temperature, which is about 2500\,K
\citep{Hebb-et-al:2009}, but to that of the lower thermosphere
which is close to the visual radius and is heated by the stellar
EUV radiation \citep{Yelle:2004,Koskinen-et-al:2012}.

We do not include stellar irradiation in the model, because this
will affect just the boundary temperature value. In our
simulation, we attempt to recover the presence of an early ingress
at the observed distance, by changing the boundary conditions to
obtain a solution where the stream from the L$_1$ point is stopped
by the stellar wind at the observed distance. On the other hand,
the influence of the boundary temperature on the solution is
rather important, because it defines the density of the
atmospheric gas at the L$_1$ point. This is a key parameter
because it is used to gather the Roche lobe overfilling and the
power of the stream from the L$_1$ point; this consequently
defines the solution. We plan to include a more accurate
description of the irradiation in a future work.

We estimated the density of the lower thermosphere considering
that the optical depth along the line of sight is $\tau=n_{\rm pl}
\times l_{\rm pl} \times k_{\rm pl}=1$. Here, the distance $l_{\rm
pl}$ corresponds to the path followed by the line of sight through
the spherical layer $[R_{\rm pl} - (R_{\rm pl}+H_{\rm pl})]$,
where $H_{\rm pl}$ is the scale height. To avoid uncertainties due
to the calculation of the opacities $k_{\rm pl}$, we assume that
the atmosphere of WASP-12b is hydrogen dominated, similarly to
what \citet{Vidal-Madjar-et-al:2003} considered for HD\,209458b,
for which the number density at the photometric radius is
$\sim\,2\times10^{10}\,\text{cm}^{-3}$
\citep{Murray-Clay-et-al:2009}. Taking into account that WASP-12b
is $\sim$\,2 times heavier and $\sim$\,1.6 times larger (radius)
than HD\,209458b, we estimated the hydrogen number density at the
photometric radius as $n_{pl}\approx
1.6\times10^{10}\,\text{cm}^{-3}$, corresponding to a density of
$\approx 2.7\times10^{-14}\,\text{g}/\text{cm}^{3}$. We defined
the obtained density at the photometric radius, and the
corresponding exobase at $1.55 R_{pl}$. For comparison,
\citet{Lai-et-al:2010} define the density at the $1$ bar level,
and their exobase at $1.59 R_{pl}$, in agreement with our
approach.

We consider a system configuration, comparable to a binary system,
consisting of the star with $M_*=1.35\,M_{\odot}$ and WASP-12b
with mass $M_{\rm pl}=1.3\times10^{-3}\,M_{\odot}$. Following the
observations, we assume that the components of this binary system,
having an orbital separation $A=4.9\,R_{\odot}$, move in a
circular orbit with a period $P_{\rm orb}= 26^h$. The linear
velocity of the planet in this system is 230\,km\,s$^{-1}$.

The flow is described by a 3D system of gravitational gas dynamic
equations closed by the equation of state of a perfect neutral
monatomic gas. In this model we neglect the non-adiabatic
processes of radiative heating and cooling. An analysis of our
solution shows that the density of the envelope is rather high, so
everywhere in the envelope the Knudsen number\footnote{The Knudsen
number is defined as: $Kn={\lambda}/H$, where $\lambda$ is the
mean free path, equal to $1/(n\times\sigma)$ where $n$ is the
number density and $\sigma$ is the cross-section,
$\sigma=10^{-15}\,cm^2$, and $H$ is a typical scale of the
solution $H=n/(dn/dr)$.} is $<1$. In the important regions of the
solution (along the streams from the L$_1$ and L$_2$ points) the
Knudsen number is always $<\,0.1$, allowing us to assume that a
gas dynamic approach is valid for the problem considered here.

In our model we use the gas dynamic equations, with the assumption
that the planetary magnetic field is small. Indeed, the
synchronization of hot-Jupiters occurs within a few of Myr
\citep[e.g.,][and references therein]{showman2002}, while the age
of the star is about 2\,Gyr \citep{Chan-et-al:2011}, hence the
planet should be tidally locked and therefore its magnetic field
be weak. There are certainly various further hypothesis for the
formation of magnetic fields in gas giants (e.g., induced fields),
but all of them do imply the presence of fields which most likely
would not protect the expanding upper atmosphere, moreover in this
case where the star is inactive \citep{Fossati-et-al:2010b}.
The presence of the large value of the magnetic field
also contradicts the observations and stellar wind interaction
models which reproduce the observed Ly$\alpha$ absorption spectra
obtained during transits of hot-Jupiters
\citep[e.g.,][]{ekenback2010,lammer2011}.

To solve the system of gas dynamic equations we use a Roe-Osher
TVD scheme of a high approximation order with the Einfeldt
modification. This numerical method allows one to study flows with
a significant density contrast. Further details of the numerical
model can be found in \citet{Boyarchuk-et-al:2002} and
\citet{Bisikalo-et-al:2003}.

Calculations have been performed in a rotating coordinate frame where the force field
can be described by the Roche potential
\begin{equation}
\Phi=-\frac{GM_*}{\sqrt{x^2+y^2+z^2}}-\frac{GM_{\rm pl}}{\sqrt{(x-A)^2+y^2+z^2}}-
\frac{1}{2}\Omega^2\left(\left(x-A\frac{M_{\rm pl}}{M_*+M_{\rm pl}}\right)^2+y^2\right) \,,
\end{equation}
here $\Omega$ is the angular velocity of the system's rotation.

The origin of the coordinate system is at the center of the star;
the $X$-axis is directed to the planet; the $Z$-axis coincides
with the rotation axis of the system and is perpendicular to the
orbital plane; the $Y$-axis finalizes the right-handed system. The
calculations have been performed in a rectangular homogeneous
grid. The size of the calculation domain is $(25 \times 20 \times
10)$\,$R_{\rm pl}$ and the grid resolution is $464 \times 363
\times 182$ cells. The accepted size of the cell of $0.05\,R_{\rm
pl}$ allows the investigation of all main flow pattern features in
the planet's vicinity.

The boundary conditions have been set as follows. The upper
atmosphere is assumed to be isothermal ($T_{\rm pl}=10^4$\,K) and
in hydrostatic equilibrium, i.e. the gas velocity in the upper
atmosphere is zero. At the level $r=R_{\rm pl}$, we set the value
of the density at the visual radius equal to
$2.7\times10^{-14}\,\text{g}/\text{cm}^{3}$. At the initial
conditions, the density of the planet upper atmosphere was defined
according to the barometric formula from the adopted boundary
conditions till the distance where the density becomes less than
the wind density. Outside this region the computational domain was
filled by gas of stellar wind. We treated the stellar wind as done
by \citet{Vidotto-et-al:2010}. The particle number density in the
stellar wind has been set equal to $5\times10^6\,\text{cm}^{-3}$
\citep{Vidotto-et-al:2010}. The stellar wind parameters proper of
WASP-12 are unknown, therefore we used that of the solar wind.
Namely, the temperature of the wind has been set equal to that of
the Sun at the corresponding distance from the star $T=10^6$\,K
\citep{Withbroe-et-al:1988}. The velocity of the wind is assumed
to be 100\,km s$^{-1}$, corresponding to the velocity of the solar
wind at the distance of WASP-12b to the host star
\citep{Withbroe-et-al:1988}. As the wind acceleration mechanism is
still an open question, in the model we imply the existence of a
wind acceleration mechanism which works similarly to light
pressure. To take this into account we modify the expression for
the Roche potential, given in Eq.~(2), as follows
$$
\Phi=-\Gamma\cdot\frac{GM_*}{\sqrt{x^2+y^2+z^2}}-\frac{GM_{pl}}{\sqrt{(x-A)^2+ y^2+
z^2}}-\frac{1}{2}\Omega^2\left(\left(x-A\frac{M_{pl}}{M_*+M_{pl}}\right)^2+y^2\right),
$$
where the $\Gamma$ coefficient is used to account for the stellar
wind acceleration and it is set to zero in the regions filled with
wind matter, and one in the rest of the calculation domain. This
allows us to avoid the non-physical wind deceleration due to
stellar gravity. The proper stellar wind velocity is subsonic with
a Mach number $M=0.85$. However, taking into account the
supersonic orbital motion of the planet ($M=1.97$) the total
velocity of the planet with respect to the stellar wind is indeed
supersonic with the rather large Mach number of $M=2.14$.

\section{The structure of the planet gaseous envelope}
\label{results}

\begin{figure}
  \includegraphics[width=160mm]{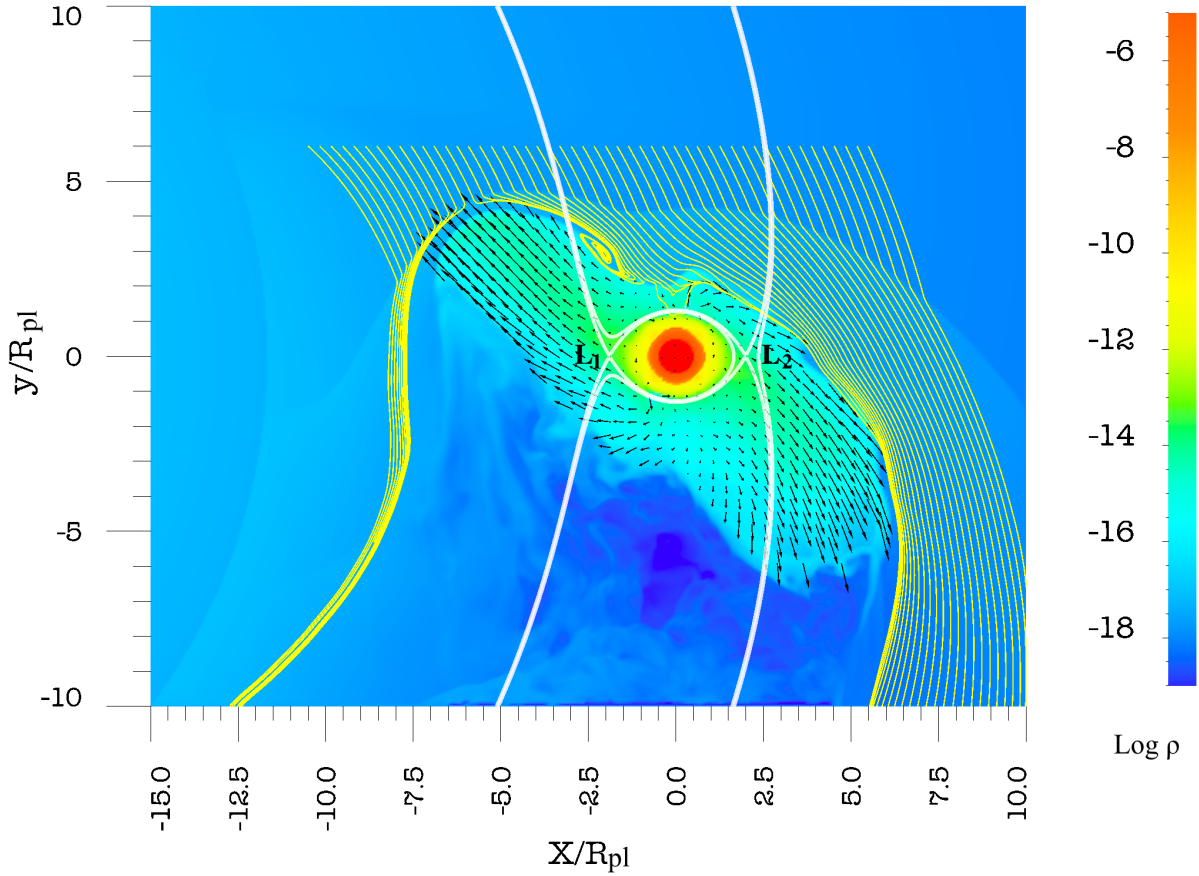}
  \caption{Density distribution and velocity vectors in the envelope of
  WASP-12b. The planet is depicted by the filled circle and it moves
  counterclockwise. The solid yellow lines denote the flow lines starting
  in the gas of the stellar wind. The white solid lines denote the Roche
  equipotentials passing through the Lagrangian points $\lp1$ and $\lp2$. }
  \label{fig1}
\end{figure}

The general morphology of the flow is shown in Fig.~\ref{fig1},
indicating the density distribution and velocity vectors in the
planet's envelope. The planet is depicted by the filled circle and
it moves counterclockwise. The solid yellow lines denote the flow
lines starting from the gas of the stellar wind. Besides, the
Roche lobe, depicted by the white solid line, and Lagrangian
points $\lp1$ and $\lp2$ are also shown.

Considering the density distribution shown in Fig.~\ref{fig1}, we
notice that the planet's envelope has a non-spherical complex
shape. In addition to the upper atmosphere itself, one can see
flows towards the $\lp1$ and $\lp2$ points. According to the
conservation of angular momentum these streams are deflected in
the direction of the planet's motion and against it, respectively.
The supersonic motion of the planet and the presence of the
stellar wind lead to the formation of a bow shock oriented
according to the total vector of the wind material velocity with
respect to the planet. The location of the shock and contact
discontinuity can be easily determined from variations in the flow
lines. It turns out that the head-on collision point is located on
the gas stream from the $\lp1$ point. The distance between the
planet center and the head-on collision point, as projected onto
the stellar limb, is $\sim$\,4.5 planet radii, in agreement with
the estimation by \citet{Lai-et-al:2010}, who derived the position
of the early-ingress at $\sim$\,3.2 planet radii, from the planet
surface.

Both shock and contact discontinuity have complex shapes. The
asymmetric shape of the planet envelope, the upper atmosphere
itself and two flows/prominences from  $\lp1$ and $\lp2$, lead to
the formation of a distinguishable double-peaked shock. The
head-on collision point is at the peak of the ``prominence'' in
the direction of $\lp1$. However, when moving closer to the
planet, we see that the shock bends due to the presence of the
planetary atmosphere, and shifts farther from the planet,  leading
to the formation of the second ``hump'' in the shock. It is
noticeable that the matter's flow, moving to the planet from the
head-on collision point, undergoes strong disturbance when it gets
into the cavity between the two humps in the shock waves. In
particular, vortices form in this cavity. This effect smooths the
contact discontinuity, mixing the matter of the stellar wind and
upper atmosphere.

The flows towards the $\lp1$ and $\lp2$ points extend far away
from the planet's Roche lobe. Formally, the planet loses its
atmosphere and, according to \eqref{f1}, the lifetime of the
atmosphere is not longer than a few years. However, in our
simulation the envelope is in a quasi-stationary regime, i.e. it
is almost enclosed. The dynamic pressure of the stellar wind and
flow, caused by the orbital motion of the planet, breaks the
propagation of the streams from the $\lp1$ and $\lp2$ points,
limiting the gaseous envelope of the planet by the bow shock and
contact discontinuity.

Let us consider the problem of the system closure in detail. From
the physical point of view, the material that has left the
planet's exosphere through the vicinity of the $\lp1$ point should
follow a certain trajectory determined by the balance of three
forces: the force of inertia, gravity forces, and the gradient of
gas pressure. We can neglect the gravity of the planet out of its
Roche lobe, hence, the gas motion can be considered as that
falling in the gravity field of the star. The stream is deflected
by the Coriolis force. However, at moderate distances from the
$\lp1$ point we can estimate the radial velocity of matter using
the difference between the values of the potential energy at the
point of our interest and the $\lp1$ point
\begin{equation}
v^2\approx \Phi({\bf L_1})-\Phi({\bf r}),\label{v2}
\end{equation}
where $\Phi({\bf r})$ is the Roche potential at the point with the
radius-vector ${\bf r}$, and ${\bf L_1}$ is the radius-vector of
the inner Lagrangian point. With the motion in the gravity field
of the star the velocity of the stream grows. However, at the same
time the wind density also grows in accordance with:
\begin{equation}
\rho_w=\rho_{w_0}\left(\dfrac{A}{|{\bf r}|}\right)^2 \,.
\label{rhow}
\end{equation}
It results that eventually the dynamic pressures of the stream
$\rho_s v^2$ and wind $\rho_w v_w^2$ become equal and the radial
motion of the stream ends. We only need to estimate at what
distance this is happening and if this distance is in agreement
with the observations.

According to ballistic analysis \citep{Lubow-Shu:75} the stream,
moving in a binary system, is deflected from the line connecting
the centers of mass of the star and planet at an angle of
$\sim\,20^\circ$. The dynamic influence of the wind should
increase this angle, and, as follows from the results of the
numerical simulations (see Fig.~\ref{fig1}), the angle grows up to
$\sim\,40^\circ$. For the location of the shock wave to correspond
to the observed one ($Y_w\approx 5\,R_{\rm pl}$ from the center)
the radial motion of the stream should stop at a distance of
$X_w=-5\,R_{\rm pl} tg 40^\circ\approx -6\,R_{\rm pl}$ from the
$\lp1$ point. We can therefore write the following equation,
setting the equilibrium between the dynamic pressures of the
stream and wind (as done by \cite{Vidotto-et-al:2010}, but for an
atmosphere with a magnetic field), and \eqref{v2} and \eqref{rhow}
\begin{equation}
\rho_{L_1}\left(\Phi({\bf L_1})-\Phi({\bf r})\right)=\rho_{w_0}\left(\dfrac{A}{|{\bf r}|}\right)^2 v_w^2 \,.
\label{balance}
\end{equation}
Solving this equation we find analytically the density at the
$\lp1$ point whose value allows us to stop the shock wave at the
observed distance from the planet. Substituting the known value of
${\bf r}$ (from the observations) into \eqref{balance}, we obtain
$\rho_{\lp1}\approx 1\times10^{-17}\,\text{g}/\text{cm}^{3}$
corresponding to a number density of $n_{\lp1}\approx
6\times10^{6}\,\text{cm}^{-3}$. The obtained value is $\sim\,18$
times less than that at the $\lp1$ point, given by the barometric
formula\footnote{When determining this value, we assumed that the
density is constant over the whole surface of the Roche lobe and
is equal to the value found for the shortest distance from the
planet to the lobe.} where
$\rho_{pl}=2.7\times10^{-14}\,\text{g}/\text{cm}^{3}$. In the gas
dynamic simulation the stream stopped at the observed distance
with the density value $\rho_{pl}\approx 1.35\times10^{-15}$,
which is $\sim\,20$ times less than the boundary density,
estimated using the planet's photometric radius. In our analysis
we omitted the effects of the stream gas heating due to the bow
shock's radiation, we can state that the analytical estimates are
in good agreement with the results of the numerical simulations.

The numerical simulations presented in this work allow us to
simulate the formation of an enclosed and steady-state gaseous
envelope surrounding ``hot Jupiters'' which exceed the planet's
Roche lobe, such as WASP-12b. We should note one more important
property of the described model of the planet's envelope.
Observations indicate \citep{Fossati-et-al:2010a} that the eclipse
in the spectral bands where the early ingress has been observed is
two times deeper than the eclipse of the planet itself (3.2\%
versus 1.7\%). The possible explanation for this effect can also
be the formation of a bow shock. Indeed, the motion of the planet
inside the stellar wind is significantly supersonic and the formed
bow shock is much hotter than the planet upper atmosphere. The
heating of the gaseous envelope including the stream from the
$\lp1$ point leads to the excitation and broadening of additional
lines in the UV bands. According to \citet{Vidal-Madjar} and
\citet{Ben-Jaffel} the depth of the transit at some spectral bands
strongly depends on its width and presence of strong spectral
lines. The larger the equivalent width of the lines in a certain
band, the deeper the transit in this band. Therefore, the
additional heating of the planet envelope can lead to the observed
effect of a stronger absorption in the NUV bands.

\section{Conclusions}
\label{concl} Using 3D numerical simulations, we investigated the
flow pattern in the vicinity of the exoplanet WASP-12b,
which is overfilling its Roche lobe. Taking into account that the
planet is tidally locked we assumed no (or negligible) surface magnetic field and consider the pure gas dynamic solution. In this work, we study the flow structure of the forming gaseous envelope surrounding the planet, without including any radiative transfer calculation. The obtained solution
depends strongly on the stellar wind parameters which we assumed close to solar.

Our results indicate that :
\begin{itemize}
\item because the planet's upper atmosphere overfills its Roche lobe,
WASP-12b's envelope has a complex non-spherical shape. In addition
to the central part of the spherical upper atmosphere two
``prominences'', directed to the $\lp1$ and $\lp2$ Lagrangian
points, develop in the system. These streams leave the planet and
are deflected in the direction of the orbital motion and against
it. Under the action of the dynamic pressure of the stellar wind
these flows slow down and, then stop at distances of $\sim\,6$ and
$\sim\,4$ planetary radii from its center, forming a stationary
envelope.

\item The planet and its envelope move in the gas of the stellar wind
with a supersonic velocity resulting in a Mach number $M=2.14$.
Thus, the dynamic pressure of the stellar wind not only works
towards the formation of the stationary envelope, but leads also
to the formation of a bow shock and contact discontinuity, which
delimit the envelope. The wave has a complex double-peaked shape.

\item The head-on collision point found in the calculations is at the
peak
of the stream from the $\lp1$ point. The distance from the planet to the
head-on collision point, as projected to the stellar limb, is of $\sim\,4.5$
planetary radii, allowing us to explain the observed extent of the early
ingress.

\item The heating of the planet envelope by the bow shock allows one to
explain the fact that the eclipse, in the spectral bands where the
early ingress has been observed, is two times deeper than the
eclipse of the planet itself.
\end{itemize}

Summarizing our results, we found that by using the full gas
dynamic simulations of the stellar wind interaction between
WASP-12b and its host star, we have been able to obtain an
alternative and self-consistent flow pattern allowing the
explanation of the existing observational data.

\section*{ACKNOWLEDGEMENTS}
This work was supported by the Basic Research Program of the
Presidium of the Russian Academy of Sciences, Russian Foundation
for Basic Research (projects 11-02-00076, 11-02-00479,
12-02-00047), Federal Targeted Program ``Science and Science
Education for Innovation in Russia 2009 -- 2013". HL and DB
acknowledge the support by the FWF NFN project S116, and the
related FWF NFN subproject S116607-N16. Finally, DB, LF, VS and HL
acknowledge support by the International Space Science Institute
in Bern, Switzerland and the ISSI team ``Characterizing Stellar-
and Exoplanetary Environments''.

\end{document}